\documentclass[prd,amsmath,twocolumn]{revtex4}

\usepackage{graphicx}
\usepackage{epstopdf}

\newcommand {\be} {\begin{equation}}
\newcommand {\ba} {\begin{eqnarray}}
\newcommand {\ee} {\end{equation}}
\newcommand {\ea} {\end{eqnarray}}

\begin{document}


\begin{flushleft}
SLAC-PUB-13513\\
\end{flushleft}

\title{Timelike Virtual Compton Scattering from Electron-Positron Radiative Annihilation} 
\author{\bf Andrei Afanasev$^a$, Stanley J. Brodsky$^b$, Carl E. Carlson$^c$, Asmita
Mukherjee$^d$}
\affiliation{
$^a$Department of Physics, Hampton University, Hampton, VA 23668, USA,
and Theory Center, Thomas Jefferson National Accelerator Facility, 12000 Jefferson Ave., Newport News, Virginia 23606, USA\\
$^b$SLAC National Accelerator Laboratory, Stanford
University, Stanford, California 94309, USA\\
$^c$Department of Physics, College of William and Mary,\\
Williamsburg, Virginia 23187, USA\\
$^d$Department of Physics,
Indian Institute of Technology, Powai, Mumbai 400076,
India   \\
}

\date{July 21, 2009}

\begin{abstract}

We propose measurements of  the deeply virtual Compton amplitude (DVCS) $\gamma^* \to H \bar H \gamma$  in the timelike  $ t = (p_{H} + p_{\bar H} )^2 > 0$ kinematic domain which is accessible at electron-positron colliders via the radiative annihilation process $e^+ e^- \to H \bar H \gamma.$  These processes allow the measurement of timelike deeply virtual Compton scattering for a variety of  $H \bar H$  hadron pairs such as  $\pi^+ \pi^- $, $K^+ K^-$, and $D \bar D$ as well as $p \bar p.$
As in the conventional spacelike DVCS, there are interfering coherent amplitudes contributing to the timelike processes involving  $C= -$ form factors. The interference between the amplitudes  measures the phase of the $C=+$ timelike DVCS amplitude relative to the phase of the timelike form factors and 
can be isolated by considering the forward-backward $e^+ \leftrightarrow e^-$ asymmetry.   The $J=0$ fixed pole contribution which arises from the local coupling of the two photons to the quark current plays a special role.  As an example we present a simple model.

\end{abstract}

\maketitle

\section{Introduction}

Deeply virtual Compton scattering  $\gamma^*(q)~ +~p(p) \to
\gamma(k)~+~ p(p^\prime),$ where the virtuality of the initial
photon $-q^2$ is large,  measures hadronic matrix elements of the current commutator $<p^\prime | [J^\mu(x) , J^\nu(0) ] |p>$ and has become a key focus in QCD, because of its direct sensitivity to fundamental hadron structure.   Assuming the handbag approximation~\cite{fac}, interactions between the virtual and real photons can be ignored, so that at  large  spacelike $q^2$  one measures matrix elements of  elementary quark  commutators 
$\sum e^2_q<p^\prime | j_q^\mu(x) , j_q^\nu(0) ] |p>$
and each
DVCS helicity amplitude factorizes as a convolution in $x$ of the
hard $\gamma^*q \rightarrow \gamma q$ Compton amplitude with a
hadronic sub-amplitude constructed from the Generalized Parton
Distributions (GPDs) $H(x,\xi,t)$, $E(x,\xi,t)$, $\tilde
H(x,\xi,t)$ and $\tilde E(x,\xi,t)$. Here $x$ is the light cone
momentum fraction of the struck quark, and the skewness $2\xi= Q^2/
(2 P \cdot q)$ measures the longitudinal momentum transfer in the
DVCS process.  


The DVCS helicity amplitudes can be constructed in the light-front formalism
from the overlap of the target hadron's light-front wave functions
\cite{overlap1,overlap2}. Since the DVCS process involves
off-forward hadronic matrix elements of light-front bilocal
currents, the overlaps are in general non-diagonal in particle
number, unlike ordinary parton distributions. Thus in the case of
GPDs, one requires not only the diagonal parton number conserving $n
\to n$ overlap of the initial and final light-front wavefunctions,
but also an off-diagonal $n+1 \to n-1$ overlap, where the parton
number is decreased by two.  Thus the GPDs measure hadron structure
at the amplitude level in contrast to the probabilistic properties
of  parton distribution functions.  In the forward limit of zero momentum transfer, the
GPDs reduce to ordinary parton distributions; on the other hand, the
integration of GPDs over $x$ at fixed skewness  $2\xi= {
Q^2/ 2 P \cdot q}$ reduces them to electromagnetic and
gravitational form factors.  One also obtains information on the orbital angular momentum carried by quarks.

The Fourier transform of the deeply virtual Compton scattering
amplitude (DVCS) with respect to the skewness parameter  $2\xi= {
Q^2/ 2 P \cdot q}$ can be used to provide an image of the target
hadron in the boost-invariant variable $\sigma$, the coordinate
conjugate to light-front time $\tau={ t+ z/ c}$~\cite{Brodsky:2006ku,Brodsky:2006in}.  The Fourier
Transform (FT) of the  GPDs with respect to the transverse momentum
transfer $\Delta_\perp$ in the idealized limit $\xi=0$ measures
the impact parameter dependent parton distributions $q(x, b_\perp)$
defined from the absolute squares of the hadron's light-front wave
functions (LFWFs) in $x$ and
impact space \cite{bur1,bur2,soper,Miller:2007uy,Carlson:2007xd}.

Virtual Compton scattering is normally measured in radiative electron-proton scattering 
$e p \to e^\prime \gamma p^\prime,$ where the  photon  virtuality  $q^2 = (p_e^\prime - p_e)^2 < 0 $ and the momentum transfer to the target proton $t = (p^\prime - p)^2 <0$  are spacelike.  The real part of the DVCS amplitude can be measured by  using the interference with the coherent Bethe-Heitler bremsstrahlung contribution  where the  real photon is emitted from the lepton in $e p \to e^\prime \gamma p^\prime.$  The interference of the DVCS amplitude and the coherent
Bethe-Heitler amplitude leads to an $e^\pm$ asymmetry which is
related to the real part of the DVCS amplitude~\cite{real}. The
imaginary part can also be accessed through various spin asymmetries~\cite{imag}.
In the deep inelastic inclusive case, the electron-positron beam asymmetry gives a three-current correlator  which is sensitive to the cube of the quark charges~\cite{Brodsky:1976fp}.

\begin{figure}[tbp]
\begin{center}

\vskip 2 mm

\includegraphics[width = 5.5 cm]{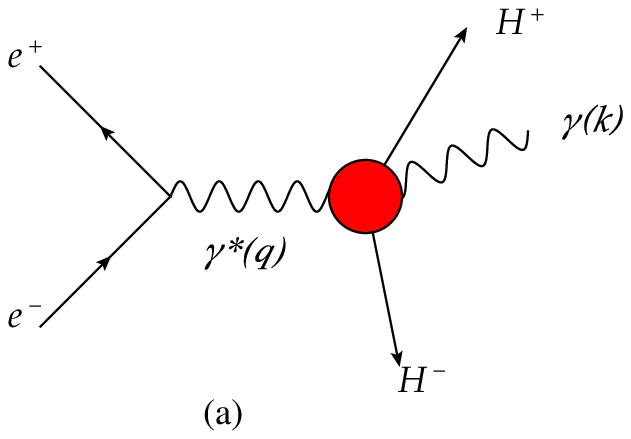}
\vskip 5 mm

\includegraphics[width = 8.4 cm]{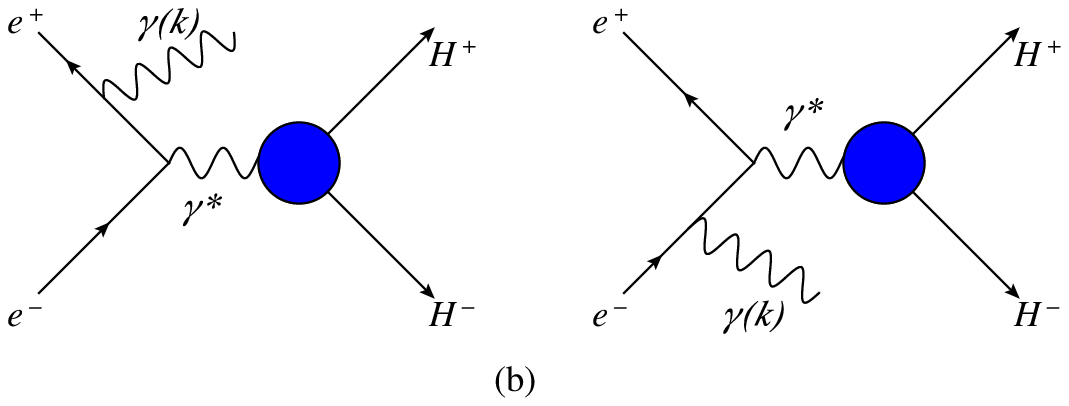}

\caption{Processes contributing to $e^+ e^- \to H^+ H^- \gamma$: (a) the generic timelike DVCS process and (b) Bethe-Heitler processes.}
\label{fig:sjb1}
\end{center}
\end{figure}

In this paper we will discuss possible measurements of  the DVCS amplitude in the timelike or 
$t >0$ kinematic domain, where $t = W^2 = (p_{H} + p_{\bar H} )^2$ is the mass of the produced hadron pair.  The processes we consider is the radiative annihilation process $e^+ e^- \to H \bar H \gamma$~\cite{Lu:2006ut}, which is accessible at electron-positron colliders and measures the timelike DVCS amplitude ${\cal M}(\gamma^*(q) \to \gamma H  \bar H )$ illustrated in Fig.~\ref{fig:sjb1}(a).  We focus on $p \bar p$ hadronic pairs, but many of the considerations apply for  a variety of $H \bar H$ hadron pairs including $\pi^+ \pi^- $, $K^+ K^-$, and $D \bar D$.
The hadronic matrix element is $C$-even since two photons attach to it.  There are also Bethe-Heitler or $C$-odd processes, Fig.~\ref{fig:sjb1}(b), that lead to the same final state.

We present sample model calculations for kinematic conditions of existing electron-positron colliders.  Relevant kinematics is chosen for tau-charm factories, s=14 GeV$^2$ (BEPCII) and B-factories,  s=112 GeV$^2$ (Babar at PEPII and Belle at KEKB) \cite{Amsler:2008zzb}.   The increased luminosity of the electron-positron colliders, such as the projected SuperB facility \cite{Biagini:2008zza}, will facilitate studies of the exclusive reactions at high transferred momenta, an example of which is considered in our paper. We note that the use of a radiative return method (see Ref.\cite{ISR} and references therein) would allow studies of the reaction of interest in a broad range of Mandelstam $s$.

\begin{figure}[tbp]
\begin{center}

\vskip 2 mm

\includegraphics[width = 5.5 cm]{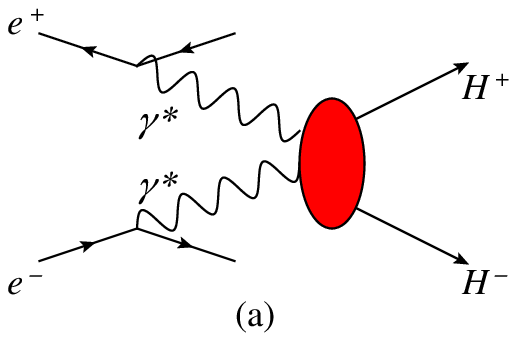}
\vskip 5 mm

\includegraphics[width = 8.4 cm]{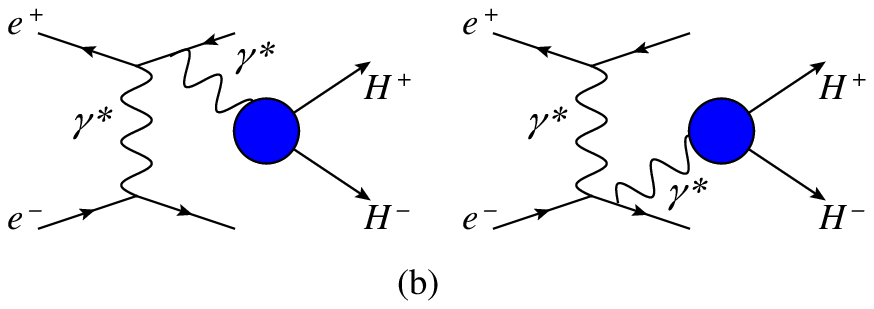}
\vskip 3 mm

\includegraphics[width = 5.5 cm]{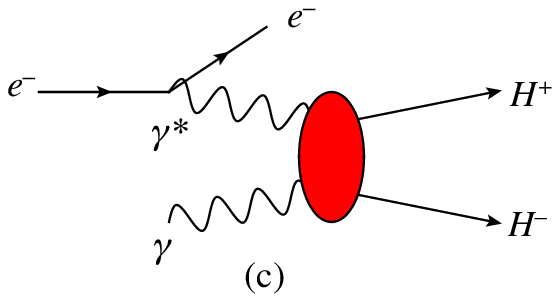}

\includegraphics[width = 5.5 cm]{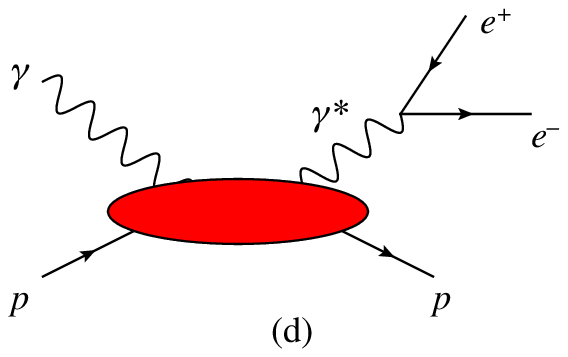}

\caption{Other processes of interest: (a) doubly virtual timelike DVCS;  (b) Bethe-Heitler processes interfering with doubly virtual timelike DVCS; (c) production of hadron pairs with one spacelike virtual photon and one real photon; and (d) a process with a timelike photon, but with spacelike momentum transfer to the hadron, which here is already present in the initial state.}
\label{fig:sjb2}
\end{center}
\end{figure}

Doubly virtual timelike DVCS~\cite{Lansberg:2006fv} is also accessible using electron-positron colliders.  One uses the process $ e^+ e^- \to  e^+ e^- H \bar H $, illustrated in Fig.~\ref{fig:sjb2}(a), to measure the amplitude ${\cal M}(\gamma^*(q) \gamma^*(q') \to  H  \bar H )$ where one or both initial photons are highly spacelike.  This process also has interfering Bethe-Heitler companion processes leading to the same final state, as in Fig.~\ref{fig:sjb2}(b).  We defer further consideration of doubly virtual processes to a future discussion. 

A process worthy of mention at this point is shown in Fig.~\ref{fig:sjb2}(c). It is production of hadron pairs from photon pairs where one photon is virtual, but in contrast to our process is spacelike rather than timelike,  and the other photon is real, but again in contrast to our process is incoming rather than outgoing.  This process is studied in Ref.~\cite{Diehl:2000uv} for outgoing pions.  These authors give the amplitude in terms of a two-hadron matrix element that they call the ``generalized distribution amplitude,'' which we might prefer to call a timelike generalized parton distribution, and whose analog we discuss in Sec.~\ref{sec:tgpd}.

The phrase ``timelike Compton scattering'' has also been applied to the process 
$\gamma p \to e^+ e^- p$~\cite{Berger:2001xd}, Fig.~\ref{fig:sjb2}(d), which is timelike in the sense that the outgoing photon is timelike.  However, this process still has spacelike momentum transfer to the nucleon, and so measures nucleon information complementary to what we are targeting here.  Also, having a pre-existing nucleon allows modeling based on known parton distribution functions, a type of modeling which is not possible here.

Measurements of the radiative annihilation process can provide valuable new information on the analytic continuation of the DVCS amplitude. The contributions to DVCS from Regge exchange in the spacelike $t < 0$ domain correspond to $C=+ $ neutral resonances in timelike domain. 
The local coupling of two photons to the fundamental quark currents of a hadron
gives a $J=0$ fixed Regge pole contribution  (\textit{i.e.}, a contribution to the amplitude that is real and constant in energy, though not necessarily constant in momentum transfer) to the Compton amplitude proportional to the charge squared of the struck quark, a  contribution 
which has no analog in hadron scattering
reactions~\cite{Brodsky:1971zh,Brodsky:1972vv,Creutz:1973zf,Brodsky:2008qu}.

In the case of ordinary spacelike DVCS this local
contribution is universal,  giving  the same contribution  for real or virtual Compton scattering for any photon virtuality and skewness at  fixed momentum transfer squared $t$. The $t$-dependence of this  $J=0$ fixed Regge pole  is
parameterized by a yet unmeasured even charge-conjugation form
factor of the target nucleon.  In the spacelike region,  
this gives an amplitude which behaves as $s^0 F^+(t)$ for $s >> -t$ corresponding to a local scalar probe. One can analytically continue the $J=0$ amplitude to a local form independent of photon virtuality at fixed $t$.  It is characterized by a complex timelike form factor $F^+(t = W^2 >0)$ dominated by scalar meson resonances.

The  $J=0$ contribution in DVCS arises 
when both photons attach locally in the same quark propagator. 
This term is the seagull interaction in the case of charged scalar quarks. The same local two-photon interaction also emerges for spin-1/2 from the usual handbag Feynman diagram for Compton scattering.  The numerator of the quark propagator $\gamma \cdot k_F + m$ appearing between the two photons in the handbag contributions to the Compton amplitude contains a specific term $ \gamma^+ \delta k^-/2$  which cancels the  $k^2_F-m^2$ Feynman denominator, leaving a local term inversely proportional to $k^+$.  This can also be identified with the instantaneous fermion exchange contribution in the light-front Hamiltonian formulation of QCD~\cite{Brodsky:1997de}.   Thus in the spin-$1/2$ case, the two-photon interaction is local in impact space and light-front time $\tau = x^+ = x^0+x^3, $  but it is nonlocal in the light-front coordinate $\sigma = x^- = x^0-x^3.$
A key feature of the $J=0$ amplitude is its independence of both $s$ and photon virtuality $q^2$ at fixed $t$. 

As already noted, for each DVCS amplitude there is an interfering coherent Bethe-Heitler amplitude where the hadronic matrix element is $C$-odd.  The interference between the amplitudes can be isolated by considering the $e^+ \leftrightarrow e^-$ asymmetry.  or the timelike cases, this does not require a separate experiment, but is equivalent to a forward-backward $e^+ \leftrightarrow e^-$ angular asymmetry in the same experiment.

The $e^+ \leftrightarrow e^-$ asymmetry measures
\ba
\label{eq:asym}
A &=& \frac { \sigma - \sigma(e^+ \leftrightarrow e^- ) }{ \sigma +  \sigma(e^+ \leftrightarrow e^- )}
		\nonumber \\
&=&  \frac{ 2\, {\rm  Re}( {\cal M}^\dagger(C=+)  \times  {\cal M} (C= - ))  }
{ |{\cal M}(C=+) |^2 + 	|{\cal M}(C= - ) |^2 }
\ea
which in turn measures the relative phase of the $C$-even DVCS amplitude and the timelike form factors.    The QED equivalents of these amplitudes,  where hadrons are replaced by muons, are useful to give a first estimate of the magnitude of the $e^+ \leftrightarrow e^-$ asymmetry.  In this paper  we obtain a simple hadronic estimate by modeling the $p \bar p$ timelike hadronic DVCS amplitude as the QED amplitude multiplied by a  $C$-even timelike form factor $R_V(\xi,W^2)$, which is related to the timelike generalized parton distribution, or the generalized distribution amplitude.  After developing the kinematics, we will calculated the $e^+ \leftrightarrow e^-$ asymmetry in a simplified model, where the input to the timelike DVCS from hadron structure appears only as a function of $ t = W^2 = (q-q')^2 $ and independent of $s$ or $q^2$.   One can say that this model simulates the C-even Compton amplitude as a $J=0$ fixed pole amplitude with Regge behavior $s^0$ at fixed $t$.   The $C=-$ amplitude is taken as the corresponding muon pair amplitude times Dirac form factor $F_1(W^2)$.
 
Of related interest is the $C=+$ form factor $R_V(t)$ from real wide-angle Compton scattering.  It is defined ratio of  the measured real Compton amplitude $M(\gamma p \to \gamma^\prime p^\prime)$ divided by  the pointlike Klein-Nishina formula.  $R_V(t)$ is measured to fall off as $1/ t^2$ at large $t$, consistent with PQCD and AdS/QCD counting rules, which in turn is consistent with what we do in our context.

Details of the calculation are given in the next section, which is divided into parts describing the kinematics, the massless pure QED limit, the continuation of the GPD analysis to the timelike region, and the actual calculation and results.   A summary and conclusions are offered in Sec.~\ref{sec:summary}.
 

\section{Cross Sections and Asymmetry}


\subsection{Kinematics}


The process is
\be
\label{eq:tdvcs}
e^+(p_{e^+}) + e^-(p_{e^-}) \to p(p_{H^+}) + \bar p(p_{H^-}) + \gamma(q')	\,,
\ee
and for comparison, we also consider the same  process with $p$ and $\bar p$ replaced by $\mu^+$ and $\mu^-$, respectively.  

The Mandelstam invariants can be defined as they are by Berends \textit{et al.}~\cite{Berends:1981rb}, namely
\begin{align}
& s = (p_{e^+} + p_{e^-})^2 ,  \  t = (p_{e^+} - p_{H})^2   , \   u = (p_{e^+} - p_{\bar H})^2   \,,  \nonumber \\
& s' = (p_{H} + p_{\bar H})^2  , \  t' = (p_{e^-} - p_{\bar H} )^2   , \  u' = (p_{e^-}  - p_{H})^2  \,.
\end{align}
Five of these variables are independent, and the sum is
\be
s+t+u+s'+t'+u' = 4 m^2  \,,
\ee
where $m$ is the mass of the hadron (or muon) in the final state, and we neglect the mass of the electron.

The cross section for process~(\ref{eq:tdvcs}) is~\cite{Lu:2006ut}
\be
d\sigma = \frac{\beta W (s-W^2) }{64 (2\pi)^5 s^2 }  \left| \cal M \right|^2  dW d\Omega^* 
d\Omega  \,,
\ee
where $| {\cal M} |^2$ is the matrix element summed over final and averaged over initial polarizations and we also use the notations
\begin{align}
s &=q^2 = Q^2  \nonumber \\
s' &= W^2  \nonumber \\
\beta &= \sqrt{1 - \frac{4m^2}{W^2}}
\end{align}
The solid angle $\Omega^*$ gives the direction of the outgoing proton or $\mu^+$ in the $p \bar p$ or $\mu^+ \mu^-$ rest frame and $\Omega$ gives the direction of the incoming electron in the $e^+ e^-$ rest frame.  We define~\cite{Lu:2006ut} the $z$ axis as the negative of the direction of the visible outgoing photon, and define the $x$-axis from the transverse direction of the proton (or $\mu^+$); see.   The angle between the proton or $\mu^+$ and the outgoing photon will be $\theta^*$, and the electron $e^-$ will enter at angles $(\theta,\phi)$ in the $e^+ e^-$ rest frame.  Thus
\ba
d\Omega^* &=&2 \pi d(\cos\theta^*)  \nonumber \\
d\Omega &=& d(\cos\theta) \, d\phi
\ea

The momenta are conveniently given using two lightlike vectors $p$ and $n$ with the property $p \cdot n = 1$.  Using these vectors,
\ba
q &=& p_{e^+} + p_{e^-} = \frac{Q}{\sqrt{2}} p + \frac{Q}{\sqrt{2}} n  \,,    \nonumber \\
\Delta &=& p_H + p_{\bar H} = \frac{Q}{\sqrt{2}} p + \frac{W^2}{Q\sqrt{2}} n  \,,  \nonumber \\
q' &=& q - \Delta = \frac{ Q^2 - W^2 }{ Q\sqrt{2} } n    \,.
\ea
In the $e^+ e^-$ rest frame, 
\ba
p = \frac{1}{ \sqrt{2} } \left( 1,0,0,1 \right)  \ , \quad   n = \frac{1}{ \sqrt{2} } \left( 1,0,0,-1 \right)  \,,
\ea
while for the $p \bar p$ (or $\mu^+ \mu^-$) CM one chooses
\ba
p = \frac{W}{ Q\sqrt{2} } \left( 1,0,0,1 \right)  \ , 
	\quad   n = \frac{Q}{ W\sqrt{2} } \left( 1,0,0,-1 \right)  \,.
\ea

\begin{figure}[htbp]
\begin{center}

\includegraphics{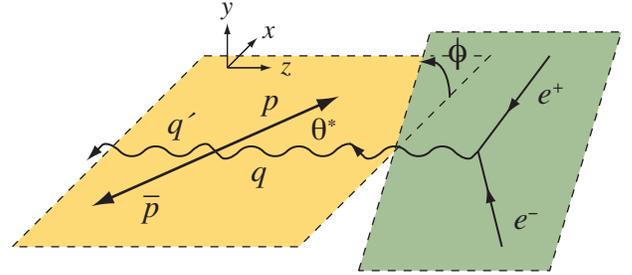}

\caption{Kinematics for radiative annihilation.  This diagram is drawn for the $p \bar p$ rest frame.  The angle $\theta$ is between the electron momentum and the $z$ axis, but in the $e^+ e^-$ rest frame.}
\label{fig:tdvcs_kinematics}
\end{center}
\end{figure}

In the $p \bar p$ (or $\mu^+ \mu^-$) rest frame the proton or $\mu^+$ momentum is
\be
p_H = \frac{W}{2}  \left(  1,  \beta \sin\theta^* , 0 , \beta \cos\theta^*  \right)
\ee
while the electron momentum in the $e^+ e^-$ CM is
\be
p_{e^-} = \frac{Q}{2}  \left(  1, \sin\theta \cos\phi , \sin\theta \sin\phi , \cos\theta  \right)
\ee
The kinematics is illustrated for the $p \bar p$ CM frame in Fig.~\ref{fig:tdvcs_kinematics}. 


The Mandelstam invariants $t$, $u$, $t'$, and $u'$ are given in terms of $Q$, $W$, and the angles by
\ba
t &=& m^2 - \frac{W^2}{4} \left( 1 - \cos\theta \right)  \left( 1 - \beta \cos\theta^* \right)
			\nonumber \\
	&& - \ \frac{Q^2}{4} \left( 1 + \cos\theta \right)  \left( 1 + \beta \cos\theta^* \right)
			\nonumber \\
	&& - \ \frac{\beta Q W}{2} \sin\theta^*  \sin\theta \cos\phi   \,,
			\nonumber \\
u &=& (\textrm{same\ as\ } t \textrm{\ but\ } \theta^* \to \pi + \theta^* )   ,
			\nonumber \\
t' &=& (\textrm{same\ as\ } t \textrm{\ but\ } \theta^* \to \pi + \theta^*,\ \theta \to \pi + \theta )  ,
			\nonumber \\
u' &=& (\textrm{same\ as\ } t \textrm{\ but\ } \theta \to \pi + \theta )  .
\ea


\subsection{The Muon Case at Zero Mass}

\begin{figure}[tbp]
\begin{center}

\vskip 2 mm

\includegraphics[width = 8.4 cm]{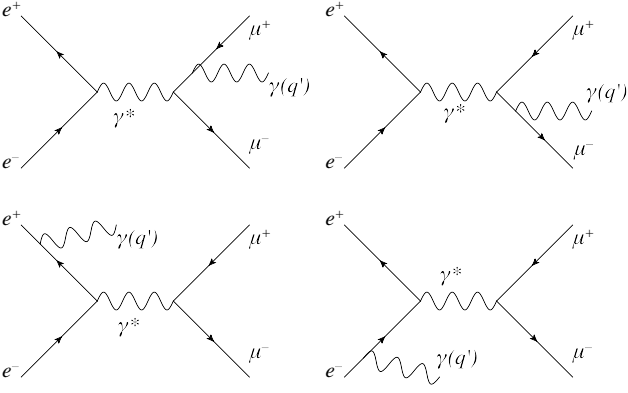}

\caption{Timelike DVCS in QED: radiative muon pair production.}
\label{fig:sjb3}
\end{center}
\end{figure}

Our calculations keep the non-zero mass of the final hadrons and are hence valid even when $s$ is close to threshold.  The analytic forms for the cross section and asymmetry for non-zero mass are rather long and we do not show them.  However, the massless limit for the pure QED calculation, Fig.~\ref{fig:sjb3} is relatively simple and valuable as a benchmark.


The matrix element for $m = 0$, summed over final and averaged over initial polarizations, splits into a factor related to the $2 \to 2$ process multiplied by a factor for the photon bremsstrahlung~\cite{Berends:1981rb},
\ba
\label{eq:2to2}
|\mathcal M|^2 = e^4 \frac{t^2 + t'^2 + u^2 +u'^2}{s s'}  S   \,,
\ea
with
\ba
\label{eq:bremss_factor}
S &=& e^2 \bigg( \frac{s}{ p_{e^+}{\cdot}q' \ p_{e^-}{\cdot}q' } 
	+ \frac{s'}{ p_H{\cdot}q' \ p_{\bar H}{\cdot}q' } 
		- \frac{t}{ p_{e^+}{\cdot}q' \ p_H{\cdot}q' } 
				\nonumber \\[1.2ex]
&&	\hskip - 6 mm - \  \frac{t'}{ p_{e^-}{\cdot}q' \ p_{\bar H}{\cdot}q' }  
	+  \frac{u}{ p_{e^+}{\cdot}q' \ p_{\bar H}{\cdot}q' } 
		+ \frac{u'}{ p_{e^-}{\cdot}q' \ p_H{\cdot}q' }    \bigg)   .
\ea
The simplicity of the above formula is a notable kinematic achievement, and more complicated earlier writings of the same quantity, for example in~\cite{Brodsky:1976fp}, can be shown after the fact to agree with it.


\begin{figure}[tbp]
\begin{center}
\includegraphics[width = 8.0 cm]{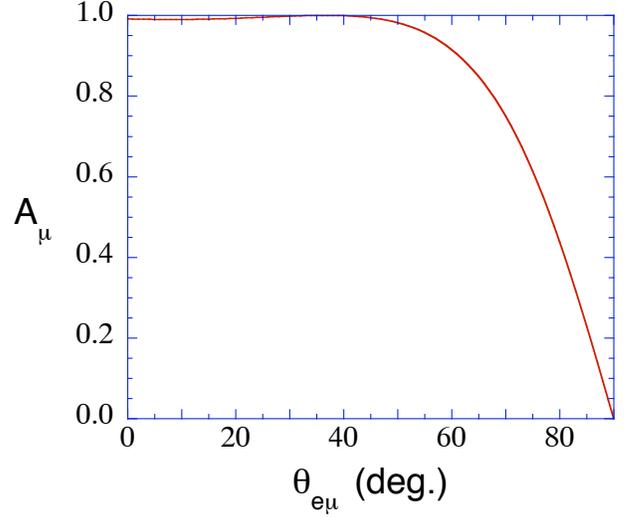}
\caption{Asymmetry where $\theta_{e\mu}$ is the angle between entering electron and exiting $\mu^+$ in the electron-positron rest frame.  Other variables are fixed as $\sqrt{s}=8$ GeV, 
$|\vec k|_\textrm{lab} = 1$ GeV, $\theta = 90^\circ$, and $\phi=0$. }
\label{fig:nomass}
\end{center}
\end{figure}


We will give a sample result for the massless limit to show that the asymmetry we wish to observe can be quite large.  The specific choices are: $\sqrt{s} = 8$ GeV, $|\vec k|_\textrm{lab} = 1$ GeV (using ``lab'' to mean the $e^+ e^-$ rest frame), all particles in the $x$-$y$ plane, and a $90^\circ$ angle between the entering electron and exiting photon in the lab.  We plot in Fig.~\ref{fig:nomass} the asymmetry
\be
A_\mu = \frac{d\sigma(\mu^+) - d\sigma(\mu^-)}{d\sigma(\mu^+) + d\sigma(\mu^-)}
\ee
versus the angle $\theta_{e\mu}$ between the electron and positive muon in the lab ({\it i.e.}, the lab analog of $\theta^*$).  This figure mimics one in~\cite{Brodsky:1976fp}, and we have obtained it here both from Eqs.~(\ref{eq:2to2}) and~(\ref{eq:bremss_factor}) above and from the massless limit of our full code.  It shows that the asymmetry is close to 100\% for a wide range of angles.


\subsection{Timelike Generalized Parton Distributions}    \label{sec:tgpd}

\begin{figure}[tbp]
\begin{center}

\includegraphics[width = 8.4 cm]{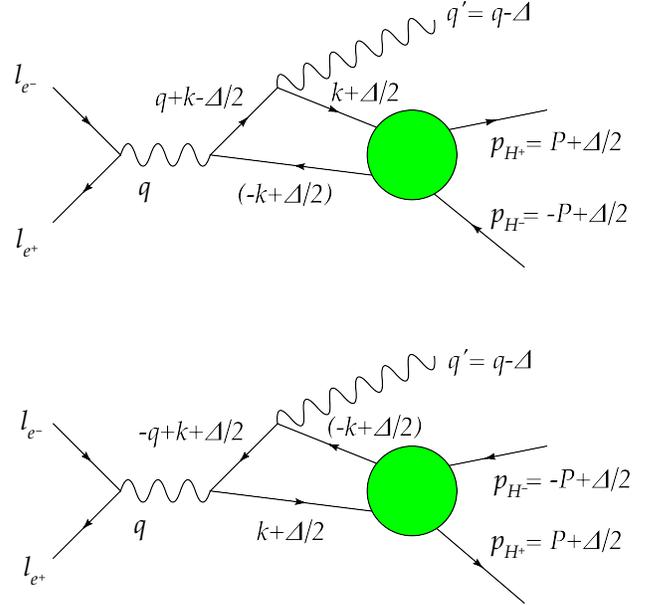}
\caption{Partonic diagrams for the case that the external photon is emitted from the hadrons.}
\label{fig:tdvcs}
\end{center}
\end{figure}

In the Bjorken limit, $Q^2 \gg W^2$ and 
$q\cdot(p_H\pm p_{\bar H}) \gg W^2$, the analysis that relates deeply virtual Compton scattering to the generalized parton distributions~\cite{fac,Ji:1996nm} can be applied in the timelike case.  The relevant diagrams at the partonic level for the $C$ even case are shown in Fig.~\ref{fig:tdvcs}.


The antihadron in the final state can be thought of as a crossing of the initial state hadron from the usual DVCS.  Changing the appropriate sign from standard DVCS definitions, one has the momentum combinations
\ba
P &=& \frac{1}{2} \left( p_H - p_{\bar H} \right)  \,,	\nonumber \\
\Delta &=& p_H + p_{\bar H} \,,
\ea
where $\Delta^2=W^2$ and $P^2 = \bar M^2 = m^2 - W^2/4 \le 0$.

The amplitude corresponding to Fig.~\ref{fig:tdvcs} is
\begin{align}
& {\mathcal M}^{\mu\nu} = - e_q^2 \int d^4z \frac{d^4k}{(2\pi)^4} e^{ikz}	\times
		\nonumber \\
& 
	\bigg[ \gamma^\mu \frac{1}{\not\! k  + \!\not\! q - \!\alpha \!\not\!\! \Delta + i\eta} \gamma^\nu
+	\gamma^\nu \frac{1}{\not\! k - \!\not\!q + (1-\alpha) \!\not\!\! \Delta + i\eta} \gamma^\mu  \bigg]_{ab}
		\nonumber \\[1.2ex]
&\times \left\langle p_H, p_{\bar H}  
	\right| T \bar\psi_a \left(-\alpha z\right) 
	\psi_b \left((1-\alpha) z \right)     \left| 0 \right\rangle	\,;
\end{align}

\noindent
$\alpha$ represents the freedom in choosing the loop momentum.
For discussing timelike generalized distribution amplitudes, or generalized distribution amplitudes~\cite{Diehl:2000uv,Diehl:1998dk}, we work in the Bjorken limit, and we choose a frame analogous to a standard choice for spacelike DVCS where the three-vectors for $P$ and $q$ are along the z-axis but in opposite directions.  We can do this with a suitable choice of the  lightlike vectors $p$ and $n$, and the momenta are expressed as
\ba
P &=& p + \frac{1}{2} \bar M^2 n \,,	
									\nonumber \\
q &=& 2 \xi p + \frac{Q^2}{4\xi} n \,,
									\nonumber \\
\Delta &=& 2\xi' \left(p - \frac{1}{2} \bar M^2 n \right) + \Delta_\perp  \,,
									\nonumber \\
k &=& x p + (p{\cdot}k) n + k_\perp  \,,
\ea
with $\xi' = \xi$ in the Bjorken limit.  In the timelike case, $\xi$ is limited in general by
\be
\frac{1}{\beta} \le \xi \le \frac{Q}{\beta W}	\,,
\ee
{\textit i.e.}, $\xi \ge 1$ in contrast to the spacelike case.
Neglecting components that do not give large contributions in the Bjorken limit, the amplitude becomes
\begin{align}
\label{eq:gpdff}
& {\mathcal M}^{\mu\nu} =  \frac{e_q^2}{2} \left( g^{\mu\nu} - p^\mu n^\nu - n^\mu p^\nu \right)
		\nonumber \\
&\quad \times		\int dx \, \left\{ \frac{1}{x+\xi+i\eta} + \frac{1}{x-\xi-i\eta} \right\}
		\nonumber \\
&\quad \times \bar u(p_H) \left[ \not\! n H^q 
	+ \frac{i}{2m} \sigma^{\alpha\beta} n_\alpha \Delta_\beta E^q  \right] v(p_{\bar H})
		\nonumber \\
&-  \frac{ i e_q^2}{2}  \varepsilon^{\mu\nu\alpha\beta} p_\alpha n_\beta
	\int dx \, \left\{ \frac{1}{x+\xi+i\eta} - \frac{1}{x-\xi-i\eta} \right\}
		\nonumber \\
&\quad \times \bar u(p_H) \left[ \not\! n \gamma^5 \tilde H^q 
	+ \frac{n \cdot \Delta}{2m} \gamma^5 \tilde E^q  \right] v(p_{\bar H})
		\nonumber \\[1.3ex]
& \equiv  - e_q^2  g^{\mu\nu}_\perp
		 \bar u(p_H) \left( \not\! n R^q_V 
	+ \frac{i}{2m} \sigma^{\alpha\beta} n_\alpha \Delta_\beta R^q_T \right) v(p_{\bar H})
		\nonumber \\
&+   i e_q^2  \varepsilon^{\mu\nu\alpha\beta} p_\alpha n_\beta
	\bar u(p_H) \left( \not\! n \gamma^5 R^q_A 
	+ \frac{n \cdot \Delta}{2m} \gamma^5 R^q_P  \right) v(p_{\bar H})	\,.
\end{align}

\noindent
We have used the definitions of the timelike analogs of the generalized parton distributions~\cite{Goeke:2001tz,Diehl:2003ny},
\begin{align}
& \int \frac{dz^-}{2\pi} e^{ix p^+ z^-}
	\left\langle p_H, p_{\bar H}  
		\right| T \bar\psi_a \left(-\frac{z}{2}  \right) 
			\psi_b \left( \frac{z}{2} \right)     
				\left| 0 \right\rangle_{z^+=z_\perp =0}
		\nonumber \\
& = \frac{1}{4} \not\! p_{ba}    \bar u(p_H) \left[ \not\! n H^q 
	+ \frac{i}{2m} \sigma^{\alpha\beta} n_\alpha \Delta_\beta E^q  \right] v(p_{\bar H})
		\nonumber \\
&\ \ +  \frac{1}{4} \left( \gamma_5 \!\not\! p \right)_{ba}
		 \bar u(p_H) \left[ \not\! n \gamma^5 \tilde H^q 
	+ \frac{n \cdot \Delta}{2m} \gamma^5 \tilde E^q  \right] v(p_{\bar H})  \,.
\end{align}
The arguments of $H^q$, $E^q$, $\tilde H^q$, and $\tilde E^q$ are $(x,\xi,W^2)$.  These arguments are standard when discussing GPDs, and the external variable $\xi$ may of course be related to $W$ and the angle $\theta^*$.  In the Bjorken limit,
\ba
\xi \approx \frac{Q^2}{4 P {\cdot} q} \approx \frac{1}{\beta  \cos\theta^* }
\,,
\ea
for $\cos\theta^*$ not too small.

For somewhat different kinematics and for the pion case, one can also find timelike analogs of GPDs defined in~\cite{Diehl:2000uv,Diehl:1998dk}.   

The form factors $R^q_V$, $R^q_T$, $R^q_A$, and $R^q_P$ are~\cite{Radyushkin:1998rt,Diehl:1998kh}
\ba
R^q_V(\xi,W^2) = \int dx \frac{x}{x^2-\xi^2 - i\eta} H^q(x,\xi,W^2)   ,
												\nonumber \\
R^q_T(\xi,W^2) = \int dx \frac{x}{x^2-\xi^2 - i\eta}  E^q(x,\xi,W^2)   ,
												\nonumber \\
R^q_A(\xi,W^2) = \int dx \frac{\xi}{x^2-\xi^2 - i\eta}  \tilde H^q(x,\xi,W^2)    ,
												\nonumber \\
R^q_P(\xi,W^2) = \int dx \frac{\xi}{x^2-\xi^2 - i\eta} \tilde E^q(x,\xi,W^2)    .
\ea


\begin{figure}[tbp]
\begin{center}

\vskip 2 mm

\includegraphics[width = 7.5 cm]{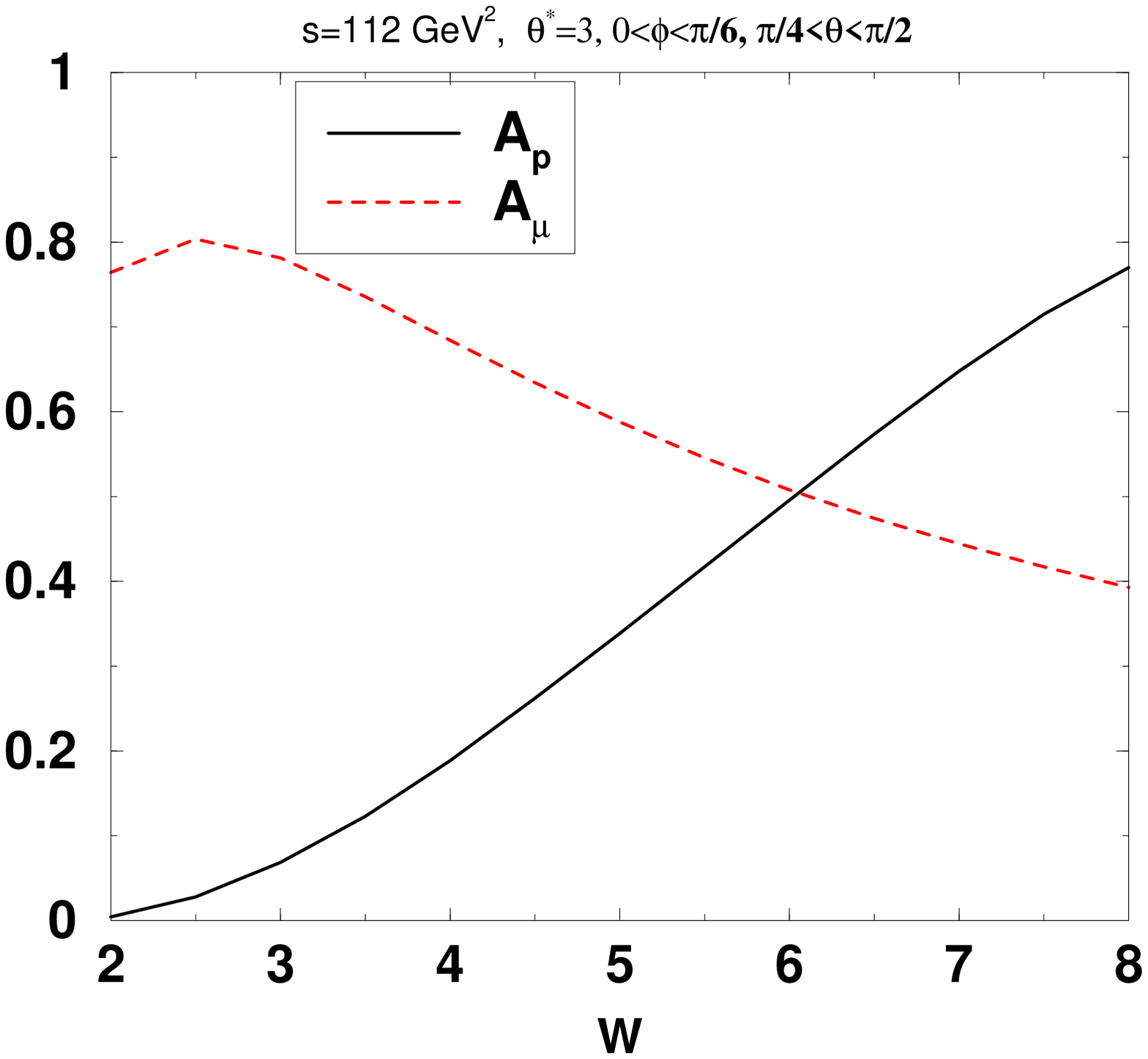}

\includegraphics[width = 7.5 cm]{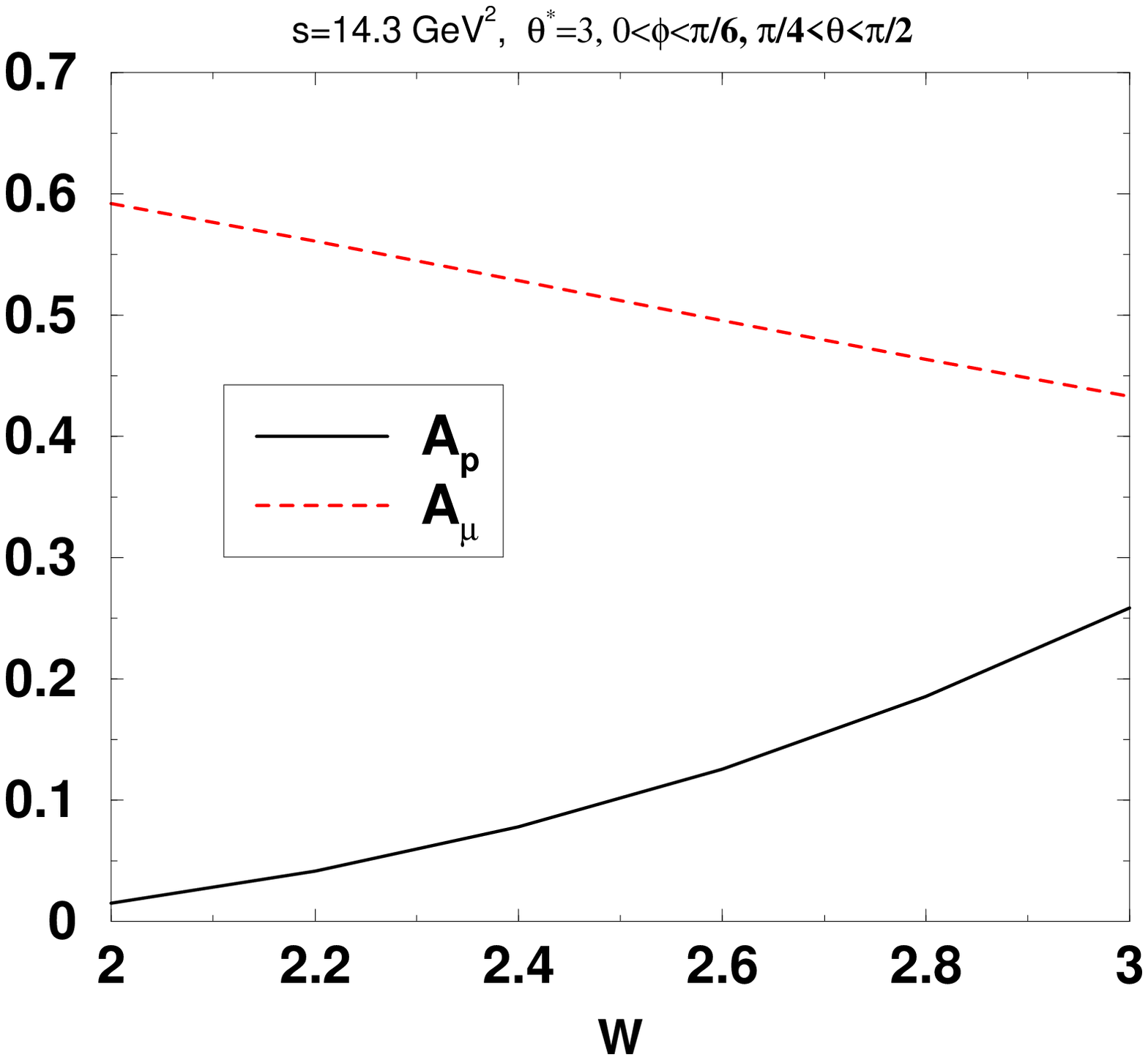}

\caption{Asymmetries for $\gamma^* \to \gamma p \bar p$ and its muonic counterpart, plotted vs. the final fermion pair invariant mass, over a range beginning close to the $p \bar p$ threshold.  The upper graph ($s = 112$ GeV$^2$) is for BELLE or Babar energies, and the lower graph ($s = 14.3$ GeV$^2$) is for BEPC II kinematics.  The angles (in radians) and angular ranges are indicated on each plot. }
\label{fig:R_V}
\end{center}
\end{figure}



\subsection{Hadronic model}


The asymmetries, Eq.~(\ref{eq:asym}), arise from interference between the $C$-odd and $C$-even amplitudes.  The hadronic vertex in the $C$-odd amplitude is the usual $\gamma p \bar p$ single photon vertex with Dirac and Pauli form factors evaluated at $W^2$, the four-momentum squared of the virtual photon for those diagrams.

For the hadronic part of the $C$-even diagrams, we have the form factors $R_{V,T,A,P}$ and here we will consider a simple model keeping only the $R_V$ form factor, given fully as
\be
R_V(\xi,t) = \sum e_q^2 R_V^q(\xi,t).
\ee
As in the spacelike case, one has the relation to the Dirac form factor,
\be
F_1(t) = \sum e_q \int dx \, H^q(x,\xi,t) \,.
\ee

\noindent
A simple ansatz is that the $t$-dependence of $R_V(\xi,t)$ and $F_1(t)$ are the same.   To get an idea of the size of the asymmetries that may exist, consider that up quarks dominate and that the explicit $x$ and $\xi$ dependent factors in the definition of $R_V^q$ roughly doubles, on average, the magnitude of the result.  Then,
\be
\label{eq:simpleapprox}
| R_V(\xi, W^2) | \approx \frac{4}{3} F_1(W^2)   \,.
\ee

We might remark that data for the spacelike case at $\xi=0$, which is obtained from wide angle real Compton scattering, show that $R_V(0,t)$ drops less rapidly with increasing $|t|$ than predicted by a model GPD based on $F_1(t)$, but is does not do so by a lot, and
\be
\frac{R_V(0,t)}{F_1(t)} \approx \frac {4}{3}	
\ee
is not bad for that situation~\cite{Danagoulian:2007gs}.  

We use the above approximation, Eq.~(\ref{eq:simpleapprox}), in the calculations that give the plots shown in Fig.~\ref{fig:R_V}.   Traces are done using FeynCalc and Mathematica, and integrations over a range of final state variables are done using Fortran and Vegas.  (Note that our modeling of the timelike GPDs is rather different 
from~\cite{Diehl:2000uv}.)

The asymmetries can be large when the kinematics are well chosen.  Fig.~\ref{fig:R_V} shows two asymmetry plots, one at $s = 112$ GeV$^2$ relevant for Belle or Babar energies and at $s = 14.3$ GeV$^2$ relevant for BEPC II energies.    The asymmetries are for cross sections integrated over a stated range of angles, and plotted versus final hadronic mass $W$.  Since the sign of the symmetry changes with $\phi$, one should not integrate over more than half the range of that angle; if desired, one can integrate over fairly broad ranges of $\theta$ and $\theta^*$.   For comparison, and to indicate the mass sensitivity for the selected $s$ and $W$, the plots also include the asymmetries expected for the purely muonic case.


\begin{figure}[bthp]
\begin{center}

\vskip 2 mm

\includegraphics[width = 7.5 cm]{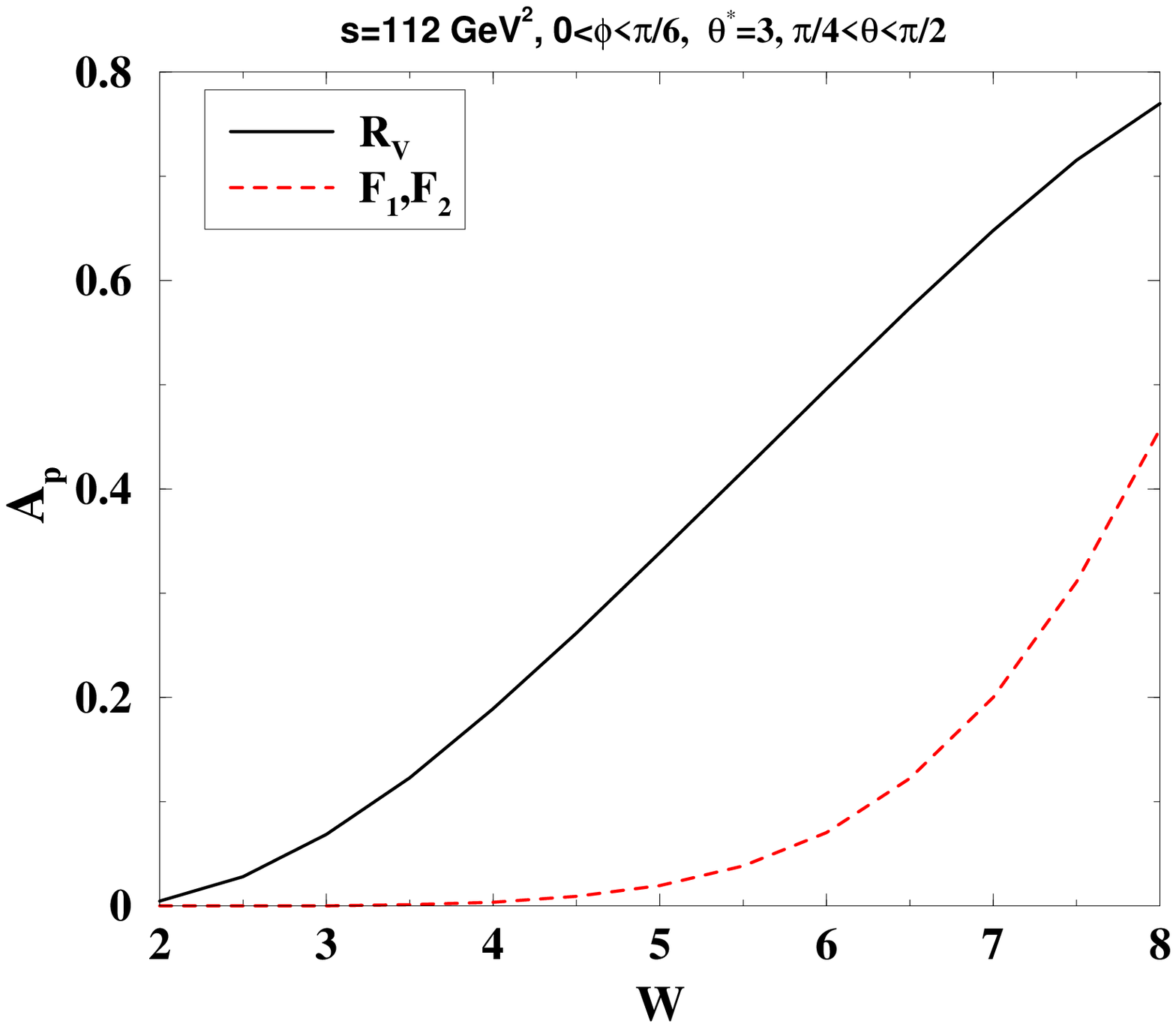}

\includegraphics[width = 7.5 cm]{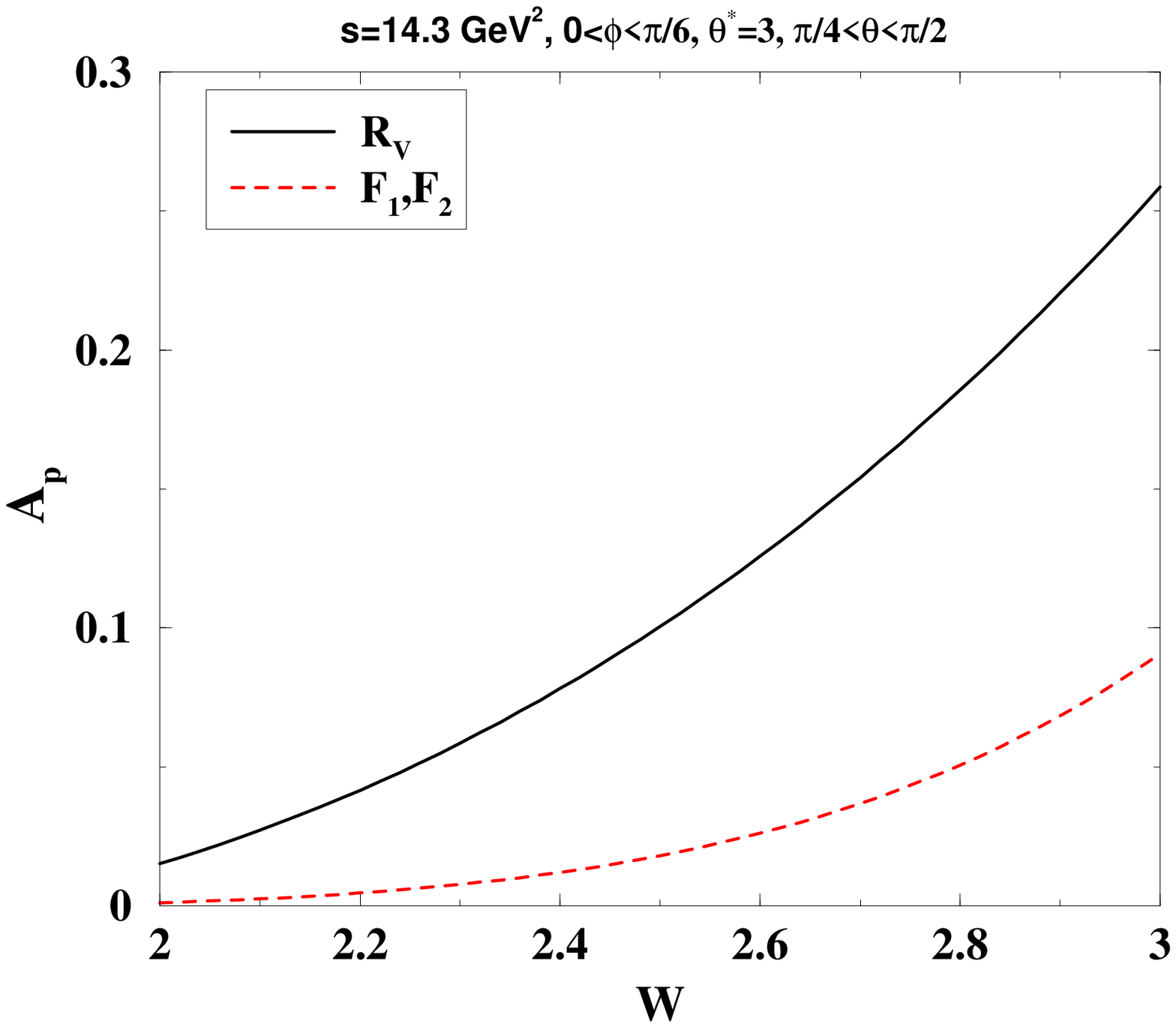}

\caption{Asymmetries for $\gamma^* \to \gamma p \bar p$ with the solid curve as in Fig.~\ref{fig:R_V}, using the $C$-even form factor $R_V(W^2)$ as described in the text, and the dashed curves calculated using a purely protonic model with form factors $F_1$ and $F_2$.}
\label{fig:F_2}
\end{center}
\end{figure}


Another comparison, shown in Fig.~\ref{fig:F_2}, follows from treating the $C$-even diagrams using only proton intermediate states, using Dirac and Pauli form factor structures at the two $\gamma p p$ vertices, and ignoring the extra form factors and extra form factor arguments that may appear when the intermediate proton is off-shell.  We keep the $F_1^2$ and $F_1 F_2$ terms, where here in the dashed curves $F_1$ and $F_2$ are 
functions of $q^2$ and $q$ is the 4 momentum of the virtual photon, and show the results in two plots that are similar to Fig.~\ref{fig:R_V} in that the energies are relevant to Belle (or Babar) and BEBC II and the cross sections are integrated over a range of angles.

The charge asymmetries are largest when the outgoing photon is at a large  
angle to the line given by electron and positron momenta in the CM.    
Conversely,  experiments that use radiative return to measure timelike  
form factors~\cite{ISR} often, though not invariably, keep the angle $ 
\theta$ below $15^\circ$ or above $165^\circ$ in order to minimize  
contributions of final state radiation.   The cross section drops  
about an order of magnitude as one changes $\theta$ from $15^\circ$ to  
the $90^\circ$ range, but the asymmetry increases, and the figure of  
merit (the cross section times asymmetry squared) stays roughly the same.

\section{Summary and Conclusions}
\label{sec:summary}

We have studied deeply virtual Compton production, $\gamma^* \to \gamma p \bar p$ in the timelike region.  The production amplitudes can be $C=+$, where both photons couple to the hadrons, or $C=-$, where only one photon couples to the hadrons.  Interference between them allows measuring one relative to the other, and can be isolated by considering forward-backward $e^+e^-$ or $p \bar p$ asymmetry.  We have used a simple model, wherein one Compton form factor, $R_V$, is kept and related to the Dirac form factor $F_1$ in a manner in agreement with data in the spacelike region.  We have found that the asymmetry is quite large and measureable.


\begin{acknowledgments}


We thank Markus Diehl, Dae-Sung Hwang, Felipe J. Llanes-Estrada, Bernard Pire, Adam P. Szczepaniak, and Werner Vogelsang for helpful discussions.  AA thanks the U.S. Department of Energy for support under U.S. DOE Contract No. DE-AC05-06OR23177.  SJB thanks the US Department of Energy for support under
grant No. DE-AC02-76SF00515.    CEC thanks the NSF for support under Grant No. PHY-0555600.  AM thanks JLab, where part of the work was done, and also William and Mary, for hospitality and support.

\end{acknowledgments}



\begin{thebibliography}{99}

\bibitem{fac} A. V. Radyushkin, Phys. Rev. {\bf D 56}, 5524 (1997); X. Ji,
J. Osborne, Phys. Rev. {\bf D 58}, 094018 (1998); J. C. Collins, A. Freund,
Phys. Rev. {\bf D 59}, 074009 (1999).

\bibitem{overlap1}  S. J. Brodsky, M. Diehl, D. S. Hwang, Nucl. Phys. {\bf B
596}, 99 (2001).

\bibitem{overlap2} M. Diehl, T. Feldmann, R. Jacob, P. Kroll, Nucl. Phys.
{\bf B 596}, 33 (2001), Erratum-ibid {\bf  605}, 647 (2001).


\bibitem{Brodsky:2006ku}
  S.~J.~Brodsky, D.~Chakrabarti, A.~Harindranath, A.~Mukherjee and J.~P.~Vary,
  Phys.\ Rev.\  D {\bf 75}, 014003 (2007)
  [arXiv:hep-ph/0611159].
\bibitem{Brodsky:2006in}
  S.~J.~Brodsky, D.~Chakrabarti, A.~Harindranath, A.~Mukherjee and J.~P.~Vary,
  Phys.\ Lett.\  B {\bf 641}, 440 (2006)
  [arXiv:hep-ph/0604262].



\bibitem{bur1} M. Burkardt, Int. Jour. Mod. Phys. {\bf A 18}, 173 (2003).

\bibitem{bur2} M. Burkardt, Phys. Rev. {\bf D 62}, 071503 (2000), Erratum-
ibid, {\bf D 66}, 119903 (2002); J. P. Ralston and B. Pire, Phys. Rev. {\bf
D 66}, 111501 (2002).

\bibitem{soper} D. E. Soper, Phys. Rev. D {\bf 15}, 1141 (1977).

\bibitem{Miller:2007uy}
  G.~A.~Miller,
  Phys.\ Rev.\ Lett.\  {\bf 99}, 112001 (2007)
  [arXiv:0705.2409 [nucl-th]];
  G.~A.~Miller, E.~Piasetzky and G.~Ron,
  Phys.\ Rev.\ Lett.\  {\bf 101}, 082002 (2008)
  [arXiv:0711.0972 [nucl-th]].

\bibitem{Carlson:2007xd}
  C.~E.~Carlson and M.~Vanderhaeghen,
  Phys.\ Rev.\ Lett.\  {\bf 100}, 032004 (2008)
  [arXiv:0710.0835 [hep-ph]]
  and arXiv:0807.4537 [hep-ph];
  Z.~Abidin and C.~E.~Carlson,
  Phys.\ Rev.\  D {\bf 78}, 071502 (2008)
  [arXiv:0808.3097 [hep-ph]].

\bibitem{real} S. J. Brodsky, F. E. Close, J. F. Gunion, Phys Rev.{\bf D
5}, 1384 (1972), {\bf D 6}, 177 (1972), {\bf D 8}, 3678 (1972).

\bibitem{imag} P. Kroll, M. Schurmann, P. A. Guichon, Nucl. Phys. {\bf A
598}, 435 (1998); M. Diehl, T. Gousset, B. Pire, J. P. Ralston, Phys. Lett.
{\bf B 411}, 193 (1997); A. V. Belitsky, D. Muller, L. Niedermeier, A.
Schafer, Phys. Lett. {\bf B 474},163,(2000).



\bibitem{Brodsky:1976fp}
  S.~J.~Brodsky, C.~E.~Carlson and R.~Suaya,
  Phys.\ Rev.\  D {\bf 14}, 2264 (1976).

\bibitem{Amsler:2008zzb}
  C.~Amsler {\it et al.}  [Particle Data Group],
  Phys.\ Lett.\  B {\bf 667}, 1 (2008).


\bibitem{Biagini:2008zza}
  M.~E.~Biagini  [SuperB Team Collaboration],
  J.\ Phys.\ Conf.\ Ser.\  {\bf 110}, 112001 (2008).

\bibitem{ISR} W.~Kluge,
  Nucl.\ Phys.\ Proc.\ Suppl.\  {\bf 181-182}, 280 (2008)
  [arXiv:0805.4708 [hep-ex]].
  
\bibitem{Ambrosino:2006gka}
  F.~Ambrosino {\it et al.},
  Eur.\ Phys.\ J.\  C {\bf 50}, 729 (2007)
  [arXiv:hep-ex/0603056],
  and DANAE Letter of Intent, at http://www.lnf.infn.it .
  
\bibitem{Lu:2006ut}
  Z.~Lu and I.~Schmidt,
  Phys.\ Rev.\  D {\bf 73}, 094021 (2006)
  [Erratum-ibid.\  D {\bf 75}, 099902 (2007)]
  [arXiv:hep-ph/0603151].

\bibitem{Lansberg:2006fv}
  J.~P.~Lansberg, B.~Pire and L.~Szymanowski,
  Phys.\ Rev.\  D {\bf 73}, 074014 (2006)
  [arXiv:hep-ph/0602195].
  
\bibitem{Diehl:2000uv}
  M.~Diehl, T.~Gousset and B.~Pire,
  Phys.\ Rev.\  D {\bf 62}, 073014 (2000)
  [arXiv:hep-ph/0003233].

\bibitem{Berger:2001xd}
  E.~R.~Berger, M.~Diehl and B.~Pire,
  Eur.\ Phys.\ J.\  C {\bf 23}, 675 (2002)
  [arXiv:hep-ph/0110062].
  
\bibitem{Brodsky:1971zh}
  S.~J.~Brodsky, F.~E.~Close and J.~F.~Gunion,
  Phys.\ Rev.\ D {\bf 5}, 1384 (1972);
  
\bibitem{Brodsky:1972vv}
  S.~J.~Brodsky, F.~E.~Close and J.~F.~Gunion,
  Phys.\ Rev.\  D {\bf 6}, 177 (1972).

\bibitem{Creutz:1973zf}
  M.~Creutz,
  Phys.\ Rev.\  D {\bf 7}, 1539 (1973).
  
\bibitem{Brodsky:2008qu}
  S.~J.~Brodsky, F.~J.~Llanes-Estrada and A.~P.~Szczepaniak,
  Phys.\ Rev.\ D {\bf 79}, 033012 (2009)
  [arXiv:0812.0395 [hep-ph]].

\bibitem{Brodsky:1997de}
  S.~J.~Brodsky, H.~C.~Pauli and S.~S.~Pinsky,
  Phys.\ Rept.\  {\bf 301}, 299 (1998)
  [arXiv:hep-ph/9705477].

\bibitem{Berends:1981rb}
  F.~A.~Berends, R.~Kleiss, P.~De Causmaecker, R.~Gastmans and T.~T.~Wu,
  Phys.\ Lett.\  B {\bf 103}, 124 (1981);
  R.~Gastmans and T.~T.~Wu,
  ``The Ubiquitous Photon: Helicity Method for QED and QCD,''
{\it  Oxford, UK: Clarendon (1990) 648 p. (International series of monographs on physics, 80)}.

\bibitem{Ji:1996nm}
  X.~D.~Ji,
  Phys.\ Rev.\  D {\bf 55}, 7114 (1997)
  [arXiv:hep-ph/9609381];
  Phys.\ Rev.\ Lett.\  {\bf 78}, 610 (1997)
  [arXiv:hep-ph/9603249].

\bibitem{Diehl:1998dk}
  M.~Diehl, T.~Gousset, B.~Pire and O.~Teryaev,
  Phys.\ Rev.\ Lett.\  {\bf 81}, 1782 (1998)
  [arXiv:hep-ph/9805380].
  
\bibitem{Goeke:2001tz}
  K.~Goeke, M.~V.~Polyakov and M.~Vanderhaeghen,
  Prog.\ Part.\ Nucl.\ Phys.\  {\bf 47}, 401 (2001)
  [arXiv:hep-ph/0106012].

\bibitem{Diehl:2003ny}
  M.~Diehl,
  Phys.\ Rept.\  {\bf 388}, 41 (2003)
  [arXiv:hep-ph/0307382].

\bibitem{Radyushkin:1998rt}
  A.~V.~Radyushkin,
  Phys.\ Rev.\  D {\bf 58}, 114008 (1998)
  [arXiv:hep-ph/9803316].

\bibitem{Diehl:1998kh}
  M.~Diehl, T.~Feldmann, R.~Jakob and P.~Kroll,
  Eur.\ Phys.\ J.\  C {\bf 8}, 409 (1999)
  [arXiv:hep-ph/9811253].

\bibitem{Danagoulian:2007gs}
  A.~Danagoulian {\it et al.}  [Hall A Collaboration],
  Phys.\ Rev.\ Lett.\  {\bf 98}, 152001 (2007)
  [arXiv:nucl-ex/0701068];
  A. Nathan, talk at the Workshop on Exclusive Reactions at High Momentum Transfer 
  (including results from JLab experiment E99-114),  May 21-24, 2007,  Jefferson Lab, Newport News, VA, USA (unpublished).


  
\end{thebibliography}
\end{document}